\documentclass[11pt]{article}
\usepackage{moriond,epsfig}

\def\Journal#1#2#3#4{{#1} {\bf #2}, #3 (#4)}

\def\PLB{{\em Phys. Lett.}  B}

\def\PRD{{\em Phys. Rev.} D}

\def\be{\begin{equation}}
\def\ee{\end{equation}}
\def\bea{\begin{eqnarray}}
\def\eea{\end{eqnarray}}

\begin{document}
\vspace*{4cm}
\title{Direct Detection of Dark Matter particles\\
with Long Range Interaction}

\author{ Paolo Panci }

\address{CP$\,^3$-Origins and DIAS,
University of Southern Denmark, \\
Odense, Denmark}

\maketitle\abstracts{We study the effect of a long--range DM-nuclei interaction occurring via the exchange of a light mediator. We consider the main direct detection experiments: DAMA, CoGeNT, CRESST, CDMS and XENON100. We find that a long--range force is a viable mechanism, which can provide full agreement between the various experiments, especially for masses of the mediator in the 10--30 MeV range and a light DM with a mass around 10 GeV. The relevant bounds on the light mediator mass and scattering cross section are then derived.}

\section{ Generalization of the pointlike cross section to long-range interactions}

A specific realization of long--range interactions between DM particles $\chi$ and target nuclei $N$ is provided by models in which ordinary photons possess a kinetic mixing $\epsilon$ with light dark photons $\phi$ (see e.g.~\cite{FPR,Holdom,Foot}). In non relativistic limit this interaction is governed by a Yukawa potential
$ V(r) = \epsilon/(4\pi) (Ze\,Z_\chi e_\chi) / r \,e^{-m_\phi r},$
whose scale is determined by the mass  of the mediator $m_\phi$.

\smallskip   
\noindent In order to have a rough comparison with the ``standard'' picture (spin independent contact interaction), the differential cross section derived from the Yukawa potential above can be cast in the following form
\begin{equation}
\frac{d\sigma(v,E_{R})}{dE_{R}}=\frac{m_N}{2 \mu_{\chi p}^2}\frac1{v^2}\,A^2 \sigma_{\phi\gamma}^p F_{\rm SI}^2(E_R) \mathcal{G}(E_R),
\end{equation}
where $\sigma_{\phi\gamma}^p=\mu_{\chi p}^2/\pi \cdot (e \, \epsilon \, Z_\chi e_\chi)^2/m_p^4$ is a normalized total cross section, $F_{\rm SI}(E_R)$ denotes the nuclear form factor and $\mu_{\chi p}$ is the DM-proton reduced mass. The function
\begin{equation}
\mathcal G(E_R)=\left(Z/A\right)^2\left[m_p^2 \,/ \left(q^2+m_\phi^2\right )\right]^2, \quad \mbox{where } q^2=2m_N E_R,
\end{equation}
measures the deviations of the allowed regions and constraints respect to the ``standard'' picture. Here the factor $(Z/A)^2$ is due to the fact that the DM only couples with protons, while the factor in the square brackets accounts for the energy dependence of the differential cross section and exhibits two limits: 
\begin{itemize}
\item[$\diamond$] Point--like limit ($q\ll m_\phi$): In this regime $\cal{G}$ is independent on $E_R$, and therefore the interaction is of contact type.

\item[$\diamond$]  Long--range limit ($q\gg m_\phi$); In this regime one has a $E_{R}^{-2}$ drop--off of the cross section (Rurtherford scattering).  Experiments with low energy thresholds and light target mass (e.g.~DAMA) are  more sensitive than the ones with high threshold and heavy targets (e.g.~XENON100). The compatibility among the experiments could therefore be improved. 
\end{itemize}

\noindent Considering typical nuclei ($m_N\sim 100$ GeV) and recoil energies (few keV)  in the range of interest of the  experiments, the transition between the two limits is obtained for $m_\phi \sim \mathcal{O} (10)$ MeV.

\section{Results}

\begin{figure}
\psfig{figure=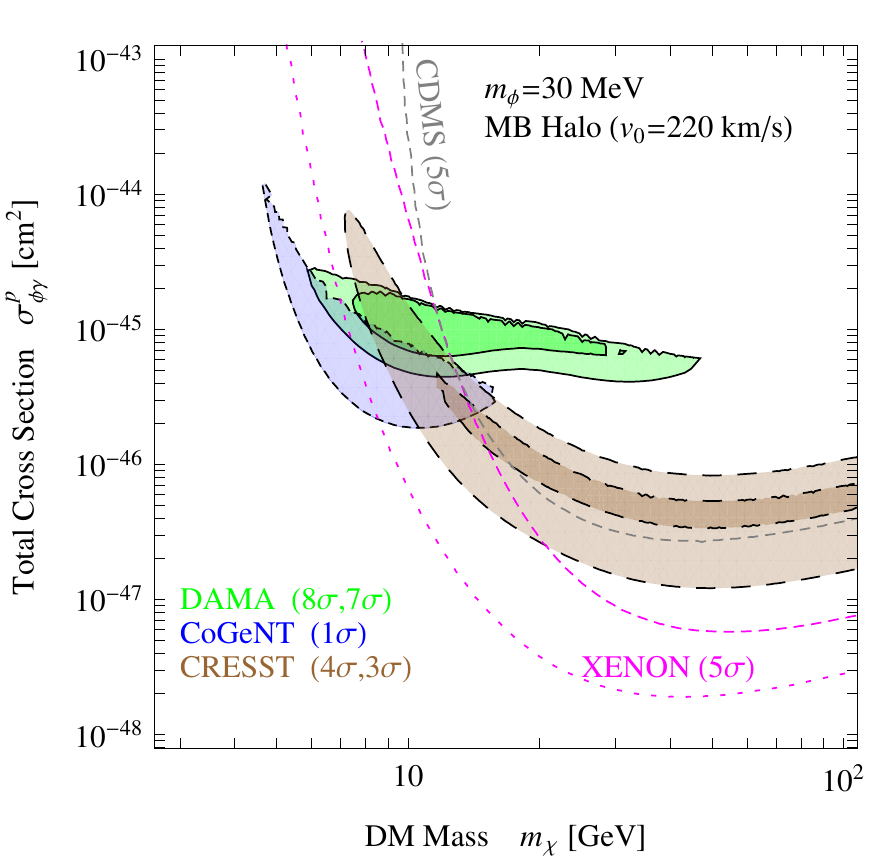, height=52.5mm}
\psfig{figure=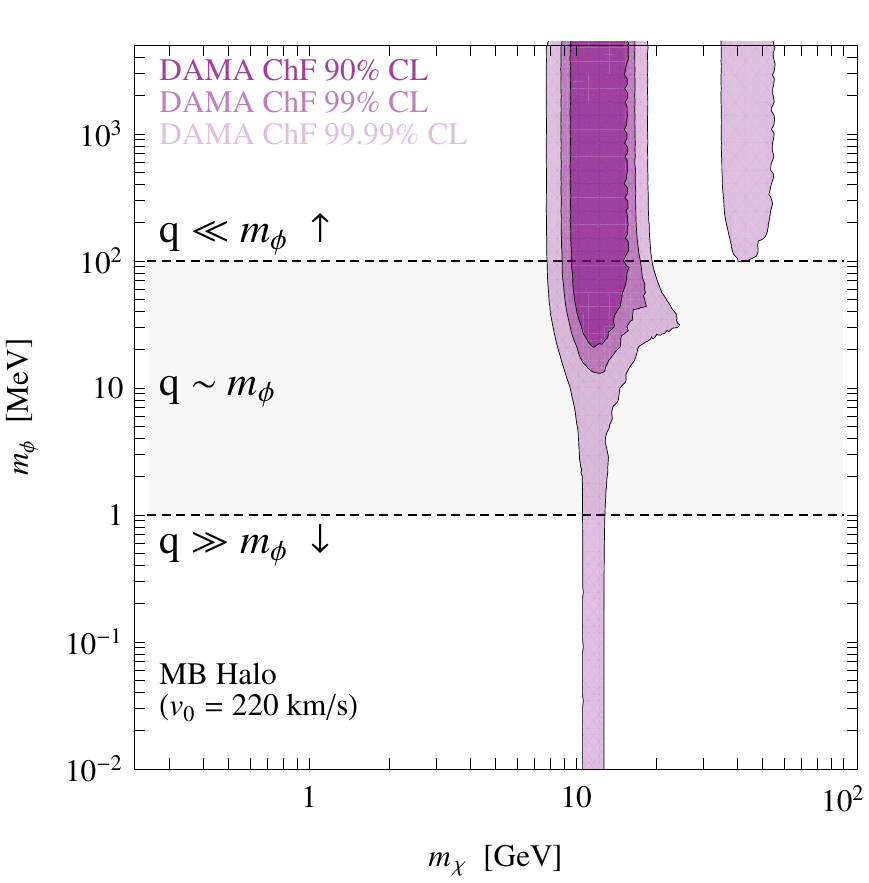, height=52.5mm}
\psfig{figure=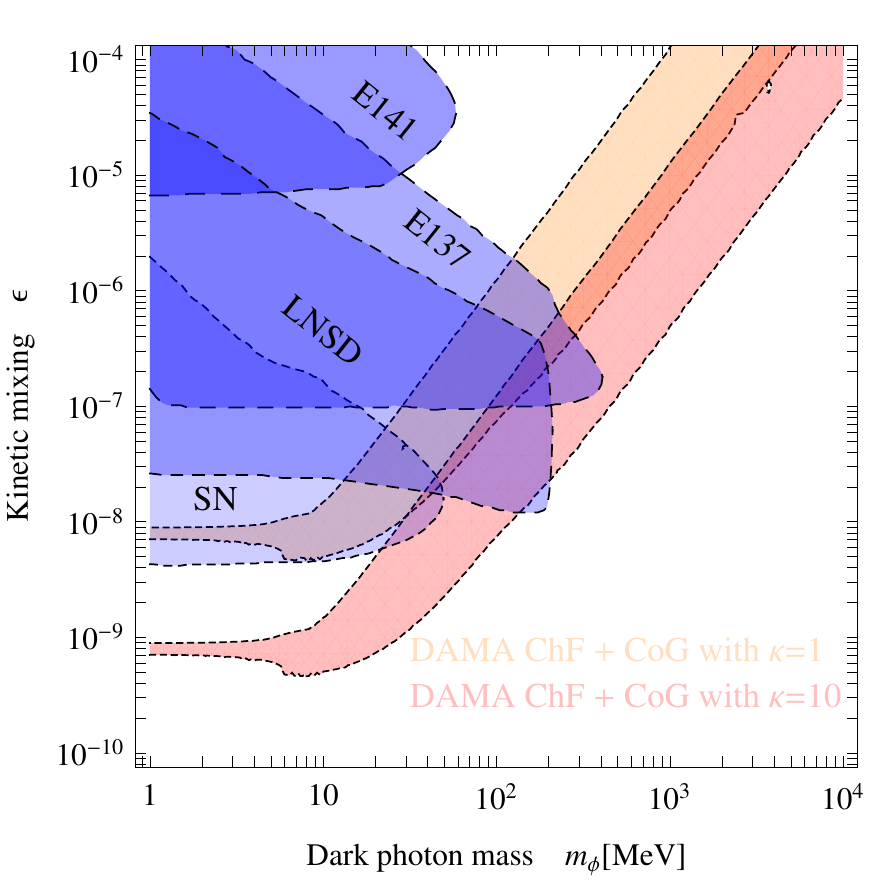, height=52.5mm}
\caption{{\emph{ left-panel}: Allowed regions compatible with DAMA (solid green regions),  CoGeNT (dotted blue regions), and CRESST (dashed brown regions) as well as the constraints coming from null results experiments (magenta and gray dashed lines for XENON100 and CDMS), considering a dark photon mass of 30 MeV; \emph{ central-panel}: Bounds on the light-mediator mass $m_\phi$ as a function of the DM mass $m_\chi$, as obtained by the analysis of the DAMA annual-modulation result; \emph{ right-panel}: Bounds and interpretation for DAMA and CoGeNT data in the $(\epsilon, m_\phi)$ plane. A summary of the constraints is discussed e.g. in~$^4$ and they are depicted as (blue) regions. }
\label{fig:1}}
\end{figure}

In the left--panel of Fig.~\ref{fig:1} we show the allowed regions and constraints coming from direct dark matter experiments, considering a dark photon mass of 30 MeV which falls in the transition between point-like and long-range interactions. One can see that:

\smallskip
$\diamond$  A large overlapping between DAMA, CoGeNT and CRESST is observed at $m_\chi \sim 10$ GeV, especially in the long--range scenario ($m_\phi\rightarrow 0$). For a conservative choice of the XENON100 constraints, the overlapping regions are allowed by all the experiments.

\smallskip
$\diamond$ Heavy targets (I for DAMA, W for CRESST), which account for the fit at large $m_\chi > 30$ GeV, reach the long--range limit earlier respect to the light ones (Na for DAMA, O and Ca for CRESST). The function $\cal G$ gets therefore suppressed and, to yield the measured event rate, an increase in $\sigma^p_{\phi \gamma}$ at high $m_\chi$ occurs. For this reason, the CRESST favorite region flattens respect to the point--like scenario. This also happens in DAMA, but the large $1/E_R^2$ enhancement of the cross section close to the low energy threshold rapidly overshoots the total rate, that is treated as a constraint; therefore, getting closer to the long--range limit, the DAMA-I region disappears.

\smallskip
$\diamond$ Approaching the long--range limit, the agreement among the various experiments increases, but the significance of the DAMA fit gets lower for the reason outlined in the previous item, this time applied to scattering on Na.

\section{Constraints}
In the central--panel of Fig.~\ref{fig:1} we show a first class of constraints in the ($m_\phi, m_\chi$) plane, coming from the fact that long--range forces rapidly overshoots the total rate being the cross section strongly enhanced. We adopt the DAMA dataset and we find that  long--range forces are only viable for light DM. Nevertheless, the best agreement is obtained for mediator masses larger than 20 MeV. A 99\% C.L. lower bound on $m_\phi$ is about 10 MeV.

\smallskip   
\noindent 
In the right--panel of Fig.~\ref{fig:1} we show instead a second class of bounds in the  ($\epsilon, m_\phi$) plane. One can see that in the ``totally symmetric'' case ($k = Z_\chi e_\chi / e = 1$), light mediators are excluded and only dark photons with $m_\phi > 100$ MeV can satisfy the constraints and provide a suitable interpretation for DAMA and CoGeNT data. However, for $k>10$ (``composite'' DM models), the whole range of light-mediator masses is allowed.  

\smallskip   
\noindent 
Another complementary class of bounds which are dependent on $Z_\chi e_\chi$ and $m_\phi$, while  independent on $\epsilon$, arises from DM self--interaction. The most famous of them comes from the Bullet Cluster~\cite{Clowe}, which points towards collisionless DM ($\sigma/m_\chi \leq 1.25$ cm$^2$/g). Considering $m_\chi \sim 10$ GeV (to fit direct detection observations) 
and again two values of the parameter $k  = 1 \, (10)$, we get that the bound on the self--interaction is exceeded for $m_\phi < 1 \, (20)$ MeV.

\end{document}